%% file: main.tex
\documentclass[aps, prx, reprint, longbibliography, nofootinbib,superscriptaddress, floatfix]{revtex4-1}

\usepackage{slashed}
\usepackage{verbatim}
\usepackage[T1]{fontenc}
\usepackage{mathbbol}
\usepackage[dvipsnames]{xcolor}
\usepackage{orcidlink}

\usepackage[normalem]{ulem}
\usepackage[english]{babel}
\usepackage{lipsum}
\usepackage{physics}
\usepackage{dcolumn}
\usepackage{tensor}
\usepackage{comment}
\usepackage{placeins}
\usepackage{graphicx,color,overpic,mathtools}
\usepackage{amsthm,amsmath,amssymb,mathrsfs}
\usepackage{braket,bm,bbm,setspace}
\usepackage{booktabs}
\usepackage{cancel}
\usepackage{float}
\usepackage{xargs}
\definecolor{myred}{RGB}{179, 27, 27}
\usepackage{hyperref}
\hypersetup{
    colorlinks=true,
    linkcolor=myred, 
    citecolor=myred, 
    urlcolor=myred  
 }

\usepackage{tikz}
\usepackage{xcolor}
\usetikzlibrary{decorations.pathmorphing}
\usetikzlibrary{decorations.markings}
\usetikzlibrary{patterns}
\usetikzlibrary{patterns.meta}
\usetikzlibrary{arrows.meta}

\newcommand{\rwh}{\tilde{h}}
\newcommand{\tth}{\breve{h}}
\newcommand{\strain}{\mathfrak{h}}
\newcommand{\bs}{{\boldsymbol{\sigma}}}

\begin{document}

\title{Second-order scattering response of a Schwarzschild black hole}

\author{Bruno Bucciotti
}
\email[]{bbucciot@asu.edu}
\affiliation{Department of Physics and Beyond: Center for Fundamental Concepts in Science,
Arizona State University, Tempe, AZ 85281, USA}
\author{Jaime Redondo-Yuste\orcidlink{0000-0003-3697-0319}}
\email[]{jredondo@princeton.edu}
\affiliation{Leinweber Forum for Theoretical Physics, Princeton University, Princeton, NJ, 08540, USA}
\author{Zhen Zhong\orcidlink{0000-0002-3138-7530}}
\email[]{zhen.zhong@uniroma1.it}
\affiliation{Dipartimento di Fisica, Sapienza Università di Roma, Piazzale Aldo Moro 5, 00185, Roma, Italy}
\affiliation{INFN, Sezione di Roma, Piazzale Aldo Moro 2, 00185, Roma, Italy
}
\author{Adrien Kuntz\orcidlink{0000-0002-4803-2998}}
\email[]{adrien.kuntz@tecnico.ulisboa.pt}
\affiliation{CENTRA, Departamento de F\'{\i}sica, Instituto Superior T\'ecnico -- IST, Universidade de Lisboa -- UL, Avenida Rovisco Pais 1, 1049-001 Lisboa, Portugal}
\author{Vitor Cardoso\orcidlink{0000-0003-0553-0433}}
\email[]{vitor.cardoso@nbi.ku.dk}

\affiliation{Center of Gravity, Niels Bohr Institute, Blegdamsvej 17, 2100 Copenhagen, Denmark}
\affiliation{CENTRA, Departamento de F\'{\i}sica, Instituto Superior T\'ecnico -- IST, Universidade de Lisboa -- UL, Avenida Rovisco Pais 1, 1049-001 Lisboa, Portugal}

\begin{abstract}
Gravitational waves couple in a nonlinear fashion in the vicinity of a black hole. Black hole perturbation theory can be readily applied to compute the magnitude of this coupling, its dependence on the parity content, frequencies, and angular structure of the incoming gravitational waves. In this work we carry out this calculation, showing how the nonlinear coupling of gravitational waves generically peaks when the driven frequency matches the oscillation frequency of the fundamental quasinormal mode of the black hole. We also demonstrate how this excitation is largest at the maximal harmonics allowed, and study its dependence on the parity content of the incoming modes. At low frequencies, the quadratic response computed here encodes the nonlinear, dynamical tidal deformability of the black hole spacetime itself. We demonstrate numerically that the quadratic black hole coupling coefficient scales quadratically with the driving frequency, and recover this scaling from the 5-point Compton graviton scattering amplitude. 
\end{abstract}

\maketitle

\section{Introduction}\label{sec:intro}

General Relativity is a nonlinear theory of the gravitational interaction. It contains gravitational-wave (GW) phenomena, which in the weak-field linear regime travel at the speed of light, and which carry two degrees of freedom, just like electromagnetic waves. However, the nonlinear character of General Relativity causes GWs to self-interact, a fascinating consequence with important implications~\cite{Wheeler:1955zz, Brill:1964zz, Isaacson:1968hbi, Isaacson:1968zza}. 

The self-interaction of GWs is imprinted in the relativistic two-body dynamics in a number of ways. Post-Newtonian calculations include a multitude of nonlinear terms~\cite{Blanchet:2013haa}, and post-Minkowskian diagrammatic calculations include graviton vertices~\cite{Damour:2016gwp, Damour:2019lcq}. Because of this ubiquity, an identification of which nonlinear terms play a role or might become important in setups other than the two-body problem is challenging.

A priori, GWs also interact with themselves during their propagation~\cite{Isaacson:1968zza}. Nevertheless, this self-interaction is suppressed when propagating through flat spacetime: a simple kinematic argument shows that it is not possible to conserve energy, helicity, and angular momentum when coupling three gravitons. Indeed, let $k_i, h_i$ be the momenta and helicities of each graviton, $i=1,2,3$. Since they are null, $k_i^2=0$. Momentum conservation means $k_i \cdot k_j=0$, so the $3$ gravitons must be collinear. Hence, angular momentum conservation requires $h_1+h_2+h_3=0$, but each helicity is $\pm 2$, so this condition can never be satisfied. A similar argument applies to photon-splitting in electrodynamics~\cite{Adler:1971}. Therefore we cannot expect -- and current observations heavily support -- that nonlinear self-interactions alter GWs in any significant way during their propagation. 

The BH ringdown, on the other hand, is an observational channel where this nonlinear coupling may be readily accessible~\cite{Berti:2025hly}. During this stage, the emission of GWs admits a particularly simple interpretation. Suffices to note that the problem is fully governed by just two parameters: the BH mass $M$, which sets the lengthscale, and its spin. The GW emission is thus governed mostly by the characteristic free oscillations of the BH, dubbed quasinormal modes (QNMs). The coupling of these QNMs, imprinted both in the GW memory~\cite{Mitman:2020pbt, Shi:2026mrr}, and directly in the ringdown~\cite{London:2014cma,Cheung:2022rbm,Mitman:2022qdl, Lagos:2024ekd, Yi:2024elj, Khera:2024bjs}, appears as a natural observational prospect. Very recently our knowledge of how QNMs couple to themselves has improved significantly, and the quadratic coupling of QNMs appears well understood~\cite{Redondo-Yuste:2023seq,Zhu:2024rej,Ma:2024qcv, Bucciotti:2024zyp, Bucciotti:2024jrv, Bourg:2024jme, Bourg:2025lpd,Khera:2024bjs}. We note that the coupling of QNMs beyond quadratic order remains a challenging open problem, see some efforts to third order in~\cite{Sberna:2021eui, Redondo-Yuste:2023ipg,Dyer:2025hdt,Emparan:2026lss}. 

While we understand how QNMs couple quadratically, we lack a systematic understanding of how GWs couple in the background of a BH spacetime. Suppose we scatter a GW with frequency $\omega$ off a BH with mass $M$. Letting $G_N$ be Newton's constant, it is well known that if $G_N M \omega \gg 1$, most of the GW will be absorbed by the BH, while if $G_N M \omega \ll 1$, the GW will only be lightly scattered. This problem is governed by the BH greybody factor, which is also imprinted in the frequency-domain GW signal~\cite{Leite:2017zyb, Oshita:2022pkc, Oshita:2023cjz, Okabayashi:2024qbz, Oshita:2024fzf, Rosato:2025ulx, Rosato:2026apq}, and in the lensing of GWs by compact objects~\cite{Futterman:1988ni,Motohashi:2021zyv,Pijnenburg:2024btj, CarrilloGonzalez:2025gqm, Chan:2025wgz, Saketh:2025cwf}. But in that same experiment, an asymptotic observer will also measure a (weak) GW with frequency $2\omega$. This is a nonlinear harmonic, generated by the self-interaction of GWs, enabled by the background curvature of spacetime. What is the amplitude of this component with frequency $2\omega$? This is the question we answer in this work. We highlight that previous numerical works~\cite{Zlochower:2003yh, Yang:2014tla, Ma:2025rnv} studied this process in the nonlinear regime (see also~\cite{Keeble:2026epf}) -- albeit a perturbative characterization, which systematically tackles all possible mode combinations, was lacking. 

In order to do so, our work extends the regularization of the second-order source of~\cite{Bucciotti:2024jrv} from QNMs to scattering data. This allows us to extend from previous results reported by some of us in~\cite{Cardoso:2026llh}, limited to a single parity sector. We find that generically the quadratic BH coupling coefficient is heavily suppressed at low frequencies, and grows near the resonance condition, when the driven frequency matches the real part of the fundamental QNM $2\omega \approx \Re[\omega_{\ell m 0}]$. At low frequencies, our computation is related to the nonlinear dynamical tidal deformability of BHs. This can be computed from the $5$-point graviton Compton scattering amplitude. To leading order, we demonstrate how this recovers the quadratic scaling with the driving frequency that we find numerically. 

\section{Set-up {\em\&} Notation}\label{sec:notation}

We study second order vacuum perturbations of a Schwarzschild spacetime,
\begin{equation}
    g_{\mu\nu}= \bar{g}_{\mu\nu} + \epsilon h^{(1)}_{\mu\nu} +\epsilon^2 h^{(2)}_{\mu\nu} \,  , 
\end{equation}
where 
\begin{equation}
    \bar{g}_{\mu\nu}dx^\mu dx^\nu = -fdt^2+f^{-1}dr^2+r^2d\Omega^2 \, , \quad f=1-\frac{2M}{r} \, ,
\end{equation}
describes a Schwarzschild BH with mass $M$. We exploit the symmetries of the background by working in the frequency domain and expanding in tensor spherical harmonics
\begin{equation}
    \begin{aligned}
        h^{(i)}_{ab} =& e^{-i\omega t} \sum_{\ell m} h^{(i)}_{ab}(r) Y^{\ell m}(\theta,\phi) \, , \\
        h^{(i)}_{aA} =& e^{-i\omega t} \sum_{\ell m} \Bigl[h^{(i)}_{a+}(r) Y_A^{\ell m}(\theta,\phi)+h^{(i)}_{a-}(r) X^{\ell m}_A(\theta,\phi)\Bigr] \, , \\
        h^{(i)}_{AB} =& e^{-i\omega t} \sum_{\ell m} \Bigl[h^{(i)}_\circ(r) q_{AB}Y^{\ell m}(\theta,\phi) + h_+^{(i)}(r) Y^{\ell m}_{AB}(\theta,\phi) \\
        &\hspace{1.7cm}+ h_-^{(i)}(r) X^{\ell m}_{AB}(\theta,\phi)\Bigr] \, . 
    \end{aligned}
\end{equation}
Above $Y^{\ell m}$, $Y^{\ell m}_A, X^{\ell m}_A$, and $Y^{\ell m}_{AB}, X^{\ell m}_{AB}$ denote scalar, vector, and tensor spherical harmonics, respectively. The indices $a,b=0,1$, denote coordinates in the Lorentzian $t,r$ plane, while $A,B=2,3$ denote angular coordinates in the $2$-sphere, and $q_{AB}$ is the usual round metric in the unit sphere. Under this decomposition, the metric perturbation decouples at each order into sectors with well-defined parity under $(\theta,\phi)\to (\pi-\theta,\pi+\phi)$ symmetry: the perturbations defined by $(h^{(i)}_{ab},h^{(i)}_{a+},h^{(i)}_\circ,h_+^{(i)})$ are even parity or \emph{polar}, whereas the perturbations defined by $(h^{(i)}_{a-}, h_-^{(i)})$ are odd parity or \emph{axial}. Each of these carries spherical harmonic indices $\ell,m$, which we omit for brevity. 

Working in perturbation theory means there is certain gauge freedom, since we can always act with a small (size $\epsilon$) coordinate transformation, leaving the background unaffected. Two gauges are particularly useful: the Regge--Wheeler (RW) gauge, which we denote with a tilde, and the transverse and traceless (TT) gauge, which we denote with a breve accent, defined by the following conditions 
\begin{equation}\label{eq:gauges}
    \begin{aligned}
        \mathrm{RW \, Gauge:}& \quad \rwh_{a+}=\rwh_+=\rwh_-=0 \, , \\
        \mathrm{TT \, Gauge:}& \quad \tth_{ab}=\mathcal{O}(r^{-2}) \, , \quad \tth_{a\pm}=\mathcal{O}(r^{-1}) \, ,  \\
        &\quad \tth_\circ = \mathcal{O}(r^0) \, , \quad \tth_\pm = \mathcal{O}(r) \, .
    \end{aligned}
\end{equation}
Notice that the TT gauge is only defined asymptotically, at large distances. On the other hand, the RW gauge is a complete gauge fixing for radiative modes, i.e., modes with $\ell \geq 2$, which are the focus of this work. 

In order to study the dynamics, we turn our attention now to the perturbed Einstein equations, which become
\begin{equation}
    \begin{aligned}
        \delta^{(1)}R_{\mu\nu}[h^{(1)}_{\mu\nu}] =& 0 \, , \\
        \delta^{(1)}R_{\mu\nu}[h^{(2)}_{\mu\nu}] =& -\delta^{(2)}R_{\mu\nu}[h^{(1)}_{\mu\nu}] \, .
    \end{aligned}
\end{equation}
At both orders, these reduce to two decoupled wave equations for certain master variables. These were first introduced by Regge and Wheeler (axial sector), and by Zerilli (polar sector), and are given, in terms of the RW gauge metric perturbation, by 
\begin{equation}
    \begin{aligned}
        \psi^{(i)}_+ &= \frac{2r}{\lambda^2}\Biggl[r^{-2}\rwh^{(i)}_\circ + \frac{2}{\Lambda}\Bigl(f^2\rwh^{(i)}_{rr} - rf(r^{-2}\rwh^{(i)}_\circ)'\Bigr)\Biggr] \, , \\
        \psi^{(i)}_- &= \frac{2r}{\mu^2}\Biggl[(\rwh^{(i)}_{t-})' +i\omega \rwh^{(i)}_{r-} - \frac{2}{r}\rwh^{(i)}_{t-}\Biggr] \, ,
    \end{aligned}
\end{equation}
where primes denote radial derivatives, and we have introduced 
\begin{equation}
\label{eq:mu_lambda_Lambda}
    \lambda^2=\ell(\ell+1) \, , \quad \mu^2=(\ell+2)(\ell-1) \, , \quad \Lambda=\mu^2+\frac{6M}{r} \, .
\end{equation}
They satisfy wave equations, which can be concisely written as 
\begin{equation}
    \frac{d^2\psi^{(i)}_\bs}{dr_*^2} + (\omega^2-V_\bs)\psi_\bs^{(i)} = S^{(i)}_{\bs}[\psi^{(i-1)},\dots,\psi^{(1)}] \, , 
\end{equation}
where $r_* = r+2M\log(r/2M-1)$ is the usual tortoise coordinate. We introduced the shorthand label $\bs=(\ell,m,\omega,\pm)$, and $V_\bs$ denotes the RW and Zerilli potentials, given by 
\begin{equation}
    \begin{aligned}
        V_+ =& \frac{f}{\Lambda^2}\Biggl[\frac{\mu^4}{r^2}\Bigl(\lambda^2+\frac{6M}{r}\Bigr) + \frac{36M^2}{r^4}\Bigl(\mu^2+\frac{2M}{r}\Bigr)\Biggr] \, , \\
        V_- =& f\Bigl(\frac{\lambda^2}{r^2} - \frac{6M}{r^3}\Bigr) \, ,
    \end{aligned}
\end{equation}
and $S^{(i)}_\bs$ is a source term that depends on the lower order perturbations. In particular, $S^{(1)}_{\bs} = 0$. The second order source can be computed by projecting onto spherical harmonics the tensor $\delta^{(2)}R_{\mu\nu}$, and can be readily found in the literature, see e.g.~\cite{Gleiser:1995gx, Ioka:2007ak, Brizuela:2006ne, Bucciotti:2024jrv}. 

\begin{figure}
    \centering
    \includegraphics[width=0.8\columnwidth]{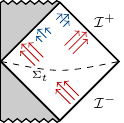}
    \caption{Penrose diagram of a Schwarzschild BH representing the scattering problem we consider. Linear fluctuations (in red) satisfy scattering boundary conditions. While they propagate in the bulk, they excite higher harmonics (in blue), which are ingoing at the horizon and outgoing at future null infinity. The dashed line shows a representative $t=\mathrm{const.}$ slice.}
    \label{fig:penrose_diagram}
\end{figure}

We will study the response of a BH to monochromatic waves. These are scattering states, where the master variables satisfy 
\begin{equation}
\label{eq:asymptotics_linear_master_scalars}
    \begin{aligned}
        \psi_\bs^{(1)} \xrightarrow{r\to2M}& e^{-i\omega r_*} \Bigl[1+\mathcal{O}(r-2M)\Bigr]  \, , \\
        \psi_\bs^{(1)} \xrightarrow{r\to\infty}& e^{-i\omega r_*}\Bigl[A^{(1)}_{\rm in}(\bs)+\mathcal{O}(r^{-1})\Bigr] 
            \\
            &+ e^{i\omega r_*}\Bigl[A^{(1)}_{\rm out}(\bs)+\mathcal{O}(r^{-1})\Bigr]  \, .
    \end{aligned}
\end{equation}
Second-order fluctuations are driven nonlinearly. Causality requires that $\psi^{(2)}_\bs$ is ingoing at the event horizon and outgoing at $\mathscr{I}^+$. This setup is shown in Fig.~\ref{fig:penrose_diagram}. 

When the linear solution $\psi_\bs^{(1)}$ is extended, the source $S_\bs^{(2)}$ typically does not decay sufficiently fast at large distances. This leads to a divergent master variable $\psi_\bs^{(2)}$. It is well known that this divergence is unphysical, and different regularization techniques have been implemented to cure it~\cite{Brizuela:2006ne, Brizuela:2009qd, Ioka:2007ak, Nakano:2007cj, Bucciotti:2024jrv}.
In essence, we will construct a \emph{regulated} master scalar $\Psi_{\bs}^{(2)}$, in analogy with eq.~\eqref{eq:asymptotics_linear_master_scalars}, with a finite amplitude at infinity
\begin{equation}
\label{eq:asymptotics_quadratic_master_scalars}
    \Psi_{\bs}^{(2)} \xrightarrow{r\to\infty} e^{i\omega r_*}\Bigl[A^{(2)}_{\rm out}(\bs)+\mathcal{O}(r^{-1})\Bigr]  \, .
\end{equation}
We tackle the problem of constructing these regulated master variables in the next section.

\section{Asymptotics and Regularization}\label{sec:asymp_reg}

We begin by revisiting the reconstruction of the GW strain at future null infinity from the RW and Zerilli master variables for linear perturbations. Next, we show how to generalize this to the second-order fluctuations, and demonstrate how to regularize the divergent behavior by defining regular master scalars. Our work extends that of Ref.~\cite{Bucciotti:2024jrv} by considering the general case of scattering fluctuations, therefore not restricted to QNMs. 

Schematically, the RW metric at first order can be reconstructed by the action of some operator $\mathcal{R}$ on the linear master functions:
\begin{equation}
    \rwh^{(1)}_{\mu\nu} = \mathcal{R}_{\mu\nu}[\psi^{(1)}] \, , 
\end{equation}
where we emphasize that $\mathcal{R}$ is a linear differential operator. Then, we can find a gauge vector $\xi^{(1)}$ such that 
\begin{equation}
    \rwh^{(1)}_{\mu\nu} + \pounds_{\xi^{(1)}}\bar{g}_{\mu\nu} = \tth_{\mu\nu} \, , 
\end{equation}
is in the TT gauge, asymptotically.
The expression of $\xi^{(1)}$ is a truncated Laurent series in $r$, which changes for asymptotically outgoing and ingoing waves. Declaring $\chi_{\rm out}=+1$ and $\chi_{\rm in}=-1$, we have for example
\begin{equation}
    (\xi^{(1)})^- = \big(\frac{\chi_{\rm in(out)}}{2r}-\frac{i}{2\omega r^2}\big)\psi^{(1)}_-|_{\rm in(out)} \, ,
\end{equation}
where $\psi_\bs^{(1)}|_{\rm in(out)}$ selects the purely ingoing/outgoing part of $\psi_\bs^{(1)}$. The full vector $\xi^{(1)}$ will be the sum of both contributions. These formulas reproduce those in \cite{Bucciotti:2024jrv} for outgoing radiation but extend them to incoming waves.

Let us define the polarization tensors 
\begin{equation}
    (e_+)^{\mu\nu} = e_\theta^\mu e_\theta^\nu - e_\phi^\mu e_\phi^\nu \, , \quad (e_\times)^{\mu\nu} = e^\mu_\theta e^\nu_\phi +e^\mu_\phi e^\nu_\theta \, , 
\end{equation}
where we introduced a tetrad 
\begin{equation}
    e^\mu_0 = \frac{\delta^\mu_0}{\sqrt{f}} \, , \quad e^\mu_r = \sqrt{f}\delta^\mu_r \ , \quad e^\mu_\theta = \frac{\delta^\mu_\theta}{r} \, , \quad e^\mu_\phi = \frac{\delta^\mu_\phi}{r\sin\theta} \, .
\end{equation}
The observable waveforms are $\strain^{(i)}_+,\strain^{(i)}_\times$ 
\begin{equation}
    \strain^{(i)}_+ = \frac{1}{2} \tth^{(i)}_{\mu\nu}(e_+)^{\mu\nu} \, , \qquad \strain^{(i)}_\times = \frac{1}{2} \tth^{(i)}_{\mu\nu}(e_\times )^{\mu\nu} \, ,
\end{equation}
and the physical GW strain amplitudes $\mathcal{A}^{(i)}_{\rm in(out)}$ are\footnote{We emphasize that spin-weight and helicity are distinct here. The combination $\strain_+-i\strain_\times$ has spin weight $-2$ for both incoming and outgoing waves. But since helicity is defined relative to the direction of propagation, the same spin-weight component corresponds to opposite helicities for the incoming and outgoing waves.}
\begin{equation}
\label{eq:strain_amplitude}
\begin{split}
    \strain^{(i)}_+(\tilde\bs) - i \strain^{(i)}_\times(\tilde\bs) = \frac{1}{r} \sum
    \big( \mathcal A^{(i)}_{\rm in}(\tilde\bs) e^{-i \omega (t+r_*)} + \\
    + \mathcal A^{(i)}_{\rm out}(\tilde\bs) e^{-i \omega (t-r_*)}\big)
    \, {}_{-2}Y^{\ell m}(\theta, \phi) \, ,
\end{split}
\end{equation}
where $\mathcal A^{(2)}_{\rm in}\equiv 0$, $\tilde\bs$ is shorthand for $(\ell,m,\omega)$, and the sum runs over all relevant quantum numbers.
After some computations, it is possible to relate $\mathcal A^{(i)}_{\rm in(out)}(\tilde\bs)$ to $A^{(i)}_{\rm in(out)}(\bs)$, defined in eqs.~\eqref{eq:asymptotics_linear_master_scalars} and \eqref{eq:asymptotics_quadratic_master_scalars}, resulting in
\begin{equation}
\label{eq:strain_map}
\begin{aligned}
    \mathcal A^{(i)}_{\rm out}(\tilde\bs)=\frac{\mu\lambda}{2}\bigl(A^{(i)}_{\rm out}(\tilde\bs,+)-iA^{(i)}_{\rm out}(\tilde\bs,-)\bigr),\\
    \mathcal A^{(1)}_{\rm in}(\tilde\bs)=\frac{\mu\lambda}{2}\bigl(A^{(1)}_{\rm in}(\tilde\bs,+)+iA^{(1)}_{\rm in}(\tilde\bs,-)\bigr) \, ,
\end{aligned}
\end{equation}
where $\mu,\,\lambda$ were defined in eq.~\eqref{eq:mu_lambda_Lambda}.

We now turn to the quadratic order. Suppose the first order perturbations are given by
\begin{equation}
    \psi^{(1)} = \psi^{(1)}_{\bs_1}+\psi^{(1)}_{\bs_2} 
    \, ,
\end{equation}
with $\bs_i = (\ell_i,m_i,\omega_i,\sigma_i)$. Whenever $\bs_1=\bs_2$ we must include an extra factor of $1/2$ in our final results, since cross terms coming from different modes carry a factor of $2$ that is absent for self-couplings. The second-order solution is $\psi^{(2)}_{\bs'}$, with $\bs' = (\ell',m', \omega', \sigma')$. The RW gauge metric can be found as 
\begin{equation}
    \rwh^{(2)}_{\mu\nu} = \mathcal{R}_{\mu\nu}[\psi_{\bs'}^{(2)}] + \mathcal{S}^{(2)}_{\mu\nu}[\psi^{(1)}_{\bs_1},\psi^{(1)}_{\bs_2}] \, ,
\end{equation}
where here we have additional contributions that depend quadratically on the two linear order master scalars $\psi^{(1)}_{\bs_1}$ and $\psi^{(1)}_{\bs_2}$.

Next, we can go to TT gauge by finding the gauge vector $\xi^{(2)}$ such that 
\begin{equation}
\label{eq:diffeo_2nd_order}
    \tth^{(2)}_{\mu\nu} = \rwh^{(2)}_{\mu\nu} + \pounds_{\xi^{(2)}}\bar{g}_{\mu\nu} + \frac{1}{2}\pounds^2_{\xi^{(1)}}\bar{g}_{\mu\nu} + \pounds_{\xi^{(1)}}\rwh^{(1)}_{\mu\nu} \, .
\end{equation}
Because $\psi_{\bs'}^{(2)}$ diverges at infinity, we now define the new master variable
\begin{align}
\begin{split}
    &\Psi_{\bs'}^{(2)} = \psi_{\bs'}^{(2)} + \Upsilon_{\bs_1,\bs_2,\bs'} [\psi^{(1)}] \, ,\\
    &\Upsilon_{\bs_1,\bs_2,\bs'} [\psi^{(1)}] = \Delta_1(r) \psi^{(1)}_{\bs_1} \psi^{(1)}_{\bs_2}
    + \Delta_2(r) f(r)\psi'^{(1)}_{\bs_1} \psi^{(1)}_{\bs_2}  \\
    &+ \Delta_3(r) f(r)\psi^{(1)}_{\bs_1} \psi'^{(1)}_{\bs_2}
    + \Delta_4(r) f(r)^2\psi'^{(1)}_{\bs_1} \psi'^{(1)}_{\bs_2} \, ,
\end{split}
\end{align}
where $\Delta_i(r)$ are truncated Laurent series in $r$ with unknown coefficients to be fixed later on.
The regulating term $\Upsilon_{\bs_1,\bs_2,\bs'} [\psi^{(1)}]$ contains all four combinations of $\psi^{(1)}_{\bs_{1,2}},\, \psi'^{(1)}_{\bs_{1,2}}$, in contrast with \cite{Bucciotti:2024jrv} that only had $\psi^{(1)}_{\bs_1} \psi^{(1)}_{\bs_2}$. Having four times the unknown parameters is what allows us to regulate all four possible combinations of in/out linear waves, to be contrasted with \cite{Bucciotti:2024jrv} where only the out--out combination appeared.

We also take
\begin{equation}
\label{eq:diffeo_2nd_ansatz}
    (\xi^{(2)})^\mu = \Gamma_1^\mu(r) \Psi_{\bs'}^{(2)} + \Gamma_2^\mu(r) \psi_{\bs_1}^{(1)} \psi^{(1)}_{\bs_2} \, ,
\end{equation}
where $\Gamma_i^\mu(r)$ are again truncated Laurent series in $r$ with unknown coefficients that depend on $\bs',\bs_1,\bs_2$. All free parameters are chosen such that:
\begin{enumerate}
    \item The variable $\Psi_{\bs'}^{(2)}$ is regular, i.e. has a finite amplitude at the horizon and infinity, and
    \item The strain is directly related to the asymptotic values of this variable. 
\end{enumerate}
These two conditions will fix the functions $\Delta_i,\,\Gamma_1^\mu,\,\Gamma_2^\mu$, rendering the problem well-defined. The first requirement is analogous to the one imposed in \cite{Bucciotti:2024jrv}, while the second one is novel: the new sources we provide are regulated so that the amplitude of the (regular) master scalars, as defined in eqs.~\eqref{eq:asymptotics_linear_master_scalars} and \eqref{eq:asymptotics_quadratic_master_scalars}, are related to the TT gauge metric perturbation eq.~\eqref{eq:gauges} in the same way at both linear and quadratic order, without additional inhomogeneous terms despite their appearance in eq.~\eqref{eq:diffeo_2nd_order},
\begin{equation}
    \begin{aligned}
        \tth^{(i)}_+|_{\rm in(out)}(\tilde\bs) &\xrightarrow{r\to\infty} r\, A^{(i)}_{\rm in(out)}(\tilde\bs,+)\, e^{i \chi_{\rm in(out)} \omega r_*} \, , \\
        \tth^{(i)}_-|_{\rm in(out)}(\tilde\bs) &\xrightarrow{r\to\infty} \chi_{\rm in(out)}\, r\, A^{(i)}_{\rm in(out)}(\tilde\bs,-)\, e^{i \chi_{\rm in(out)} \omega r_*} \, .
    \end{aligned}
\end{equation}
With the ansatz~(\ref{eq:diffeo_2nd_ansatz}), all four combinations of in/outgoing linear modes can be smoothly brought to TT gauge simultaneously. We note that at the next perturbative order, we expect divergences to appear, since an infinite monochromatic driving is unphysical. 

\section{Second-Order Response} \label{sec:second_order_response}

The coupling of two GWs with given frequencies and angular momenta on top of a background symmetric under time translations and rotations is subject to the usual selection rules, familiar from quantum mechanics. In particular:
\begin{equation}
    \begin{aligned}
        \textbf{(i): } \quad &m' = m_1+m_2 \, , \\
        \textbf{(ii): } \quad &\omega' = \omega_1+\omega_2 \, , \\
        \textbf{(iii): } \quad &\max(2,|m'|,|\ell_1-\ell_2|)\leq \ell' \leq \ell_1+\ell_2 \, , \\
        \textbf{(iv): } \quad &\sigma_1\sigma_2\sigma' = (-1)^{\ell_1+\ell_2+\ell'} \, .
    \end{aligned}
\end{equation}
The third selection rule should be extended to allow for $\ell'=0,1$. However, these are nonradiative modes and require a separate treatment. 

Our goal is to compute the second order response of the BH. This is given by the measured GW strain at future null infinity of the second-order fluctuations, compared to that of the incoming first-order GWs, 
\begin{equation} \label{eq:Q_def}
    \mathcal{Q}_{\tilde\bs_1\times \tilde\bs_2\to \tilde\bs'} = \frac{G_N M \mathcal A^{(2)}_{\rm out}(\tilde \bs')}{\mathcal A^{(1)}_{\rm in}(\tilde\bs_1)\mathcal A^{(1)}_{\rm in}(\tilde\bs_2)} \, . 
\end{equation}
The second-order coupling coefficient of the BH $\mathcal{Q}$ is a fundamental property that depends only on (i) the geometry of the BH, and (ii) the nonlinear content of Einstein's equations. Therefore, characterizing this quantity, and connecting it with GW observables appears a unique probe simultaneously of the BH hypothesis and of the nonlinear, dynamical regime of General Relativity. 

Notice that we could also define the relative amplitude of the \emph{outgoing} modes:
\begin{equation}
    \mathcal{Q}^{\rm out}_{\tilde\bs_1\times \tilde\bs_2\to \tilde\bs'} = \frac{G_N M \mathcal A^{(2)}_{\rm out}(\tilde \bs')}{\mathcal A^{(1)}_{\rm out}(\tilde\bs_1)\mathcal A^{(1)}_{\rm out}(\tilde\bs_2)} \, . 
\end{equation}
Doing so allows us to directly relate the value of the analytical continuation of $\mathcal{Q}^{\rm out}$ to complex frequencies with the excitation ratio of quadratic QNMs, well-known now in the literature~\cite{Berti:2025hly}. However, from a scattering process perspective, we believe the ratio between out and ingoing modes is better motivated. The relation between these two is, nevertheless, quite simple
\begin{equation}
\label{eq:Q_in-Q_out}
    \mathcal{Q}^{\rm out}_{\tilde\bs_1\times \tilde\bs_2\to \tilde\bs'} = \frac{\mathcal{Q}_{\tilde\bs_1\times \tilde\bs_2\to \tilde\bs'}}{\mathcal{R}_{\tilde\bs_1}\mathcal{R}_{\tilde\bs_2}} \, , 
\end{equation}
with $\mathcal{R}_{\tilde\bs}$ the BH reflectivity. 

The incoming GW strain $\strain_{\sigma_i}^{(1)}|_{\rm in}$ is directly determined by the incoming amplitudes. In order to determine the outgoing, second order strain $\strain^{(2)}_{\sigma}$ we refine the algorithm outlined in~\cite{Cardoso:2026llh}. We solve the wave equation for the regularized master variable $\Psi^{(2)}$, which is 
\begin{equation}
    \frac{d^2\Psi^{(2)}_{\bs'}}{dr_*^2}+(\omega'^2-V_{\bs'})\Psi^{(2)}_{\bs'} = \mathfrak{S}_{\bs'}[\psi^{(1)}] \, 
\end{equation}
where the regularized source is obtained as 
\begin{equation}
    \begin{aligned}
        \mathfrak{S}_{\bs'}[\psi^{(1)}] &= S^{(2)}_{\bs'}[\psi^{(1)}] \\
        &+ \Bigl[\frac{d^2}{dr_*^2}+(\omega'^2-V_{\bs'})\Bigr]\Upsilon_{\bs_1,\bs_2,\bs'} [\psi^{(1)}]
        \, .
    \end{aligned}
\end{equation}
We solve this equation using variation of parameters. At large distances, it follows that 
\begin{equation}
    \Psi_{\bs'} = \frac{e^{i\omega' r_*}}{2i\omega' A_{\rm in}(\bs')} \int dr'_* \psi_{\rm in, \bs'}(r') \mathfrak{S}_{\bs'}(r') 
    , 
\end{equation}
with $\psi_{\rm in, \bs'}$ the homogeneous solution which is ingoing at the event horizon. Although this is a highly oscillatory integral, the regularization carried out above ensures that it decays sufficiently fast as $r_*\to \pm \infty$, yielding a finite result. 

We have implemented the evaluation of the homogeneous solutions, construction of the regularized source, and evaluation of the oscillatory integral in an open source \texttt{Julia} repository \texttt{BlackHoleNonlinearities.jl}, accessible via~\cite{BlackHoleNonlinearities}. The homogeneous differential equations are integrated with adaptive, high-order Runge-Kutta methods, with boundary conditions near the horizon provided by a $10$th order Frobenius expansion of the equations. Outgoing amplitudes are extracted by matching the solution to a truncated asymptotic series expansion up to $30$th order. The integral is evaluated using adaptive Gauss-Kronrod quadrature, and we later use Richardson extrapolation to improve the convergence and extrapolate our finite-domain evaluation of the integral to $r\to\infty$. Our results converge with sub-percent accuracy as long as the integration domain is sufficiently large, as we demonstrate in the Appendix. 

\subsection{Time-domain cross-check}

We have checked these frequency-domain results with an independent time-domain calculation, using two regularized second-order masters: $\Psi^{(2)}_{\bs'}$ with the source $\mathfrak S$, and
the master $\Psi^{(2)}_{{\rm BMT},\bs'}$ of Ref.~\cite{Brizuela:2009qd} (BMT) with its regularized source
$\mathfrak S_{\rm BMT}$, described below.
In both cases the first- and second-order master equations are evolved in Kerr--Schild coordinates $(t_{\mathrm{KS}},r)$, with $t_{\mathrm{KS}}=t+r_*-r$ the Kerr--Schild time, using a Chebyshev
multidomain spectral method for the spatial discretization and 4th-order Runge--Kutta for the time integration.
The inner boundary lies inside the horizon, at $r<2M$, and needs no boundary condition; the outer
boundary lies at $r\geq300M$, where we impose the boundary condition on the incoming characteristic and leave the
outgoing one free.

The first-order initial data are the stationary scattering state of each parent at its frequency, $\propto e^{-i\omega t}$,
and its incident wave is imposed at the outer boundary, so the first-order field remains in this state up to
numerical errors.

On the scattering
state, $\mathfrak S$ oscillates at the sum frequency $\omega'=\omega_1+\omega_2$.
We start
$\Psi^{(2)}_{\bs'}$ from zero and extract it once the startup transient has decayed at the extraction radii. At the outer boundary, which lies at a finite radius, the second-order field is not simply an outgoing wave: its
large-$r$ expansion is the outgoing wave plus the
asymptotic particular solution $\Psi^{(2)}_{{\rm p},\bs'}$ driven by $\mathfrak S$, with no free homogeneous
term, from which we construct the
boundary condition.

The master $\Psi^{(2)}_{{\rm BMT},\bs'}$ regularizes the same bare master with a different quadratic term, $Q_{\rm reg}$, and obeys, in Schwarzschild
coordinates,
\begin{equation}
(-\partial_t^2+\partial_{r_*}^2-V_{\bs'})\,\Psi^{(2)}_{{\rm BMT},\bs'}=\mathfrak S_{\rm BMT},
\end{equation}
with $\mathfrak S_{\rm BMT}=f\,{}^{(2)}\mathcal S^{\rm reg}_\Psi$,
where ${}^{(2)}\mathcal S^{\rm reg}_\Psi$ is BMT's regularized source of that master. Unlike $\mathfrak S$,
which is written in the frequency domain, $\mathfrak S_{\rm BMT}$ is written in the time domain and holds for any first-order solution; in Kerr--Schild coordinates it is finite at the horizon. Because BMT constructed it for outgoing radiation, it does not decay at large $r$ on the scattering state,
which also contains incoming waves.
Hence a
second-order field started from zero would be far from its final state at large $r$. We therefore start it from the
asymptotic particular solution $\Psi^{(2)}_{{\rm BMT,p},\bs'}$ multiplied by a smooth function that is zero for $r<60M$ and one for
$r>120M$. The outer boundary condition is constructed in the same way, from this particular solution.

To obtain $\mathcal Q$ from the BMT evolution, we convert its amplitude. On the monochromatic scattering
state, where $\Upsilon$ is defined,
the source identity fixes the two masters up to a homogeneous solution:
\begin{equation}
\begin{split}
&(\partial_{r_*}^2+\omega'^2-V_{\bs'})\bigl(\Psi^{(2)}_{\bs'}-\eta\,\Psi^{(2)}_{{\rm BMT},\bs'}-\delta\Upsilon\bigr)=0,\\
&\delta\Upsilon\equiv\Upsilon-\eta\,Q_{\rm reg},
\end{split}
\end{equation}
where the constant $\eta$ accounts for their different normalizations and depends on the channel. Far from the black
hole each parent contains both incoming and outgoing waves, so
\begin{equation}
\begin{split}
\delta\Upsilon\sim{}&\delta\Upsilon_{\rm out}A^{(1)}_{\rm out}(\bs_1)A^{(1)}_{\rm out}(\bs_2)\,e^{i\omega'r_*}\\
&+\delta\Upsilon_{\rm in}A^{(1)}_{\rm in}(\bs_1)A^{(1)}_{\rm in}(\bs_2)\,e^{-i\omega'r_*}\\
&+\Psi^{(2)}_{{\rm p},\bs'}-\eta\,\Psi^{(2)}_{{\rm BMT,p},\bs'}.
\end{split}
\end{equation}
At large $r$ each master is an outgoing wave plus its particular solution, so the $e^{-i\omega'r_*}$ term must be removed;
we do so with $\psi_{{\rm in},\bs'}$, which also carries an outgoing part:
\begin{equation}\label{eq:bmt_transport}
\begin{split}
A^{(2)}_{\rm out}(\bs')={}&\eta\,A^{(2)}_{\rm out,BMT}(\bs')
+\delta\Upsilon_{\rm out}A^{(1)}_{\rm out}(\bs_1)A^{(1)}_{\rm out}(\bs_2)\\
&-\frac{A_{\rm out}(\bs')}{A_{\rm in}(\bs')}\,\delta\Upsilon_{\rm in}A^{(1)}_{\rm in}(\bs_1)A^{(1)}_{\rm in}(\bs_2),
\end{split}
\end{equation}
where $\psi_{{\rm in},\bs'}\to A_{\rm in}(\bs')\,e^{-i\omega'r_*}+A_{\rm out}(\bs')\,e^{i\omega'r_*}$ at large $r$.

We extract $\mathcal Q$ in a late-time window. At radii between $50M$ and $240M$ we take the component of
each field and its radial derivative at its frequency; at second order we first subtract the asymptotic particular solution
driven by its source, as at
the outer boundary. The remainder is fitted
to the incoming and outgoing homogeneous solutions. We convert the BMT amplitude with Eq.~\eqref{eq:bmt_transport}, map the master amplitudes to strain
amplitudes with Eq.~\eqref{eq:strain_map}, and obtain $\mathcal Q$ from Eq.~\eqref{eq:Q_def}. Our time-domain results agree with the frequency-domain calculation to sub-percent accuracy. An example of this agreement is shown in Fig.~\ref{fig:selfcoup}.

\section{Results}

We compute the quadratic BH coupling coefficient $\mathcal{Q}_{\tilde\bs_1\times\tilde\bs_2\to\tilde\bs'}$, defined in Eq.~\eqref{eq:Q_def}, for scattering states of arbitrary driving frequencies $\omega_1,\omega_2$, across all radiative $(\ell,m)$ and parity combinations allowed by the selection rules of Sec.~\ref{sec:second_order_response}. The response depends on five independent angular labels, two parent frequencies, and the allowed parity content. Rather than attempting an exhaustive scan of this (large) parameter space, we focus on four regimes of interest: the low-frequency limit, the resonant behavior near QNM frequencies, the angular redistribution of the quadratic response, and the dependence on the frequency whenever $\omega_1 \neq \omega_2$. Our code is available in~\cite{BlackHoleNonlinearities}.

\subsection{Low-frequency regime}

We begin by examining the low-frequency limit, $G_N \omega' M \ll 1$. As we will discuss later, we reproduce qualitatively this limit by matching to the calculation of a $5$-point Compton amplitude. The main challenge in the low-frequency regime is that the integration domain has to be sufficiently large to resolve the long wavelength of the impinging GWs -- typically of size $r_{\rm max} \sim 10/\omega$. This somewhat limits our ability to probe the low-frequency regime below $\omega' \lesssim 10^{-3}$. 

For simplicity, we focus on the case where $\omega_1=\omega_2$, and consider a variety of modes that can couple with the leading $(\ell_1,m_1)=(2,2)$ mode, in different parity sectors. In all cases, we find that the magnitude of the BH response is quadratic in the frequency, $|\mathcal{Q}|\sim (G_N M \omega')^2$, confirming earlier results~\cite{Cardoso:2026llh}. We come back to this point later, as we will show that this scaling can be justified from scattering amplitudes. In particular, the $5$-point Compton scattering amplitude for gravitons~\cite{Bjerrum-Bohr:2023jau} precisely predicts a scaling like $(G_N M \omega')^2$ for $\mathcal{Q}$. We leave open for future work extending this to higher orders, and fixing the numerical coefficient as a function of the angular momenta. The phase only varies slowly with $\omega'$ -- this encodes the conservative response, which we expect enters at a higher perturbative order in $G_N M\omega$. 

\begin{figure}
    \centering
    \includegraphics[width=\columnwidth]{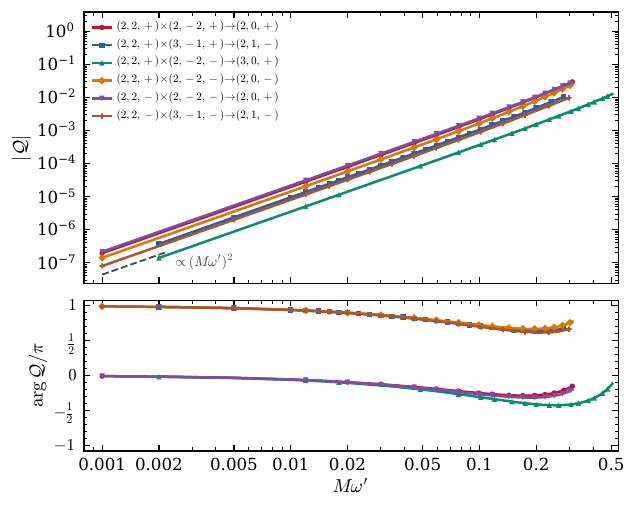}
    \caption{Absolute value (top) and phase (bottom) of the quadratic BH coupling coefficient $\mathcal{Q}$ as a function of the driven frequency $\omega'$, for different channels coupling with the $(2,2)$ mode. In all cases the scaling with the frequency is quadratic, while the phase varies only slowly with $\omega'$. }
    \label{fig:low_freq}
\end{figure}

\subsection{Near-Resonance regime}
 The characteristic oscillation modes of a BH -- its QNMs -- have complex frequencies. Therefore, there are no exact resonances when driving the BH with a \emph{real} frequency. Still, one can consider a quasi-resonance which takes place when the driven frequency $\omega' \approx \Re \omega_{\ell' m' 0}$. Early results identified an amplification of the BH response near this driving frequency~\cite{Cardoso:2026llh}. Now we confirm that this amplification holds for generic channel combinations, and for all possible parities. 

\begin{figure}
     \centering
     \includegraphics[width=\columnwidth]{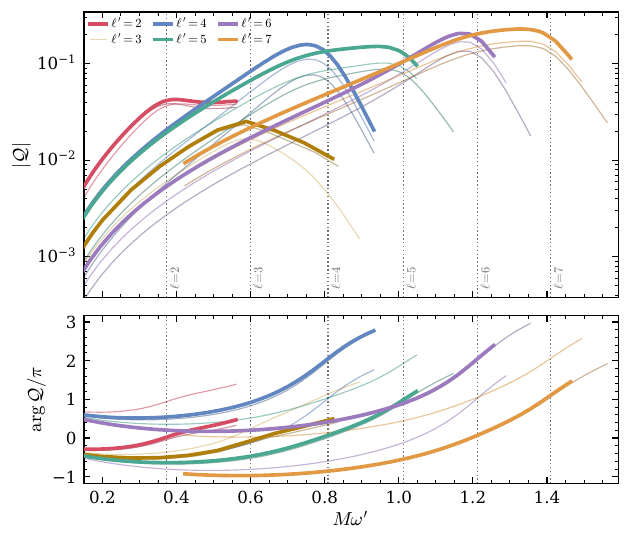}
     \caption{Absolute value (top) and phase (bottom) of the quadratic coupling coefficient $\mathcal{Q}$ as a function of the driving frequency $\omega'$, for different values of the outgoing angular number $\ell'$, as indicated in the legend. The different lines within the same color correspond to different channels that lead to the same value of $\ell'$: they all behave qualitatively similarly, showing a clear peak when $\omega'$ matches the real part of the QNM frequencies -- indicated as a dashed, vertical line.}
     \label{fig:resonance}
 \end{figure}

 Fig.~\ref{fig:resonance} shows the absolute value and phase of $\mathcal{Q}$ as a function of the driving frequency. Once again, we restrict ourselves to the case where $\omega_1=\omega_2$ --- see below for the effect of unequal frequencies. The different colors correspond to different values of the driven angular harmonic $\ell'$. Recall that for Schwarzschild BHs, the azimuthal number $m$ does not affect the QNM frequencies: $\omega_{\ell' m' 0} = \omega_{\ell' 0 0}$. The different shades of the same colors correspond to different angular and parity combinations that result in the same value of $\ell'$. In all cases, we see that the lines are peaked near the dashed vertical lines, which mark the real part of the corresponding QNM frequency. Moreover, the peak amplitude generally grows with $\ell'$, more strongly for the purely polar channels with maximal azimuthal number $|m_i|=\ell_i$. It is interesting to note that while there is a clear local maximum near the QNM frequency in the absolute value of $|\mathcal{Q}|$, there is no smoking gun of this resonance in the phase.

\subsection{Eikonal Limit}

In order to study the behavior at higher frequencies, we consider the self-coupling $(\ell,\ell,\omega)\times (\ell,\ell,\omega) \to (2\ell,2\ell,2\omega)$ as a function of $\ell$. This allows us to probe the eikonal regime as $\ell \to \infty$. Notice that carrying the numerical integration at high driving frequencies, $G_N M\omega'\to\infty$ becomes computationally expensive, as one needs to resolve a large number of oscillations of a source that only decays polynomially. Working on a compactified domain, or further regularizing the source so that it decays faster, could allow us to probe the higher frequency limit more accurately. 

To avoid clutter, we study only the purely even combination $++\to +$. We do not find significant qualitative difference with respect to other parity combinations. Our results show that, when normalizing the frequency by the real part of the relevant QNM frequency, all curves peak right before the resonant condition. More interestingly, as $\ell$ grows, the curves tend to overlap. This suggests that the large $\ell$ limit may be universal, and admit a simple expression. In particular, based on our results we conjecture that at the lightring frequency $|\mathcal{Q}(\omega' = 2\ell \Omega_{\rm LR})|$ may be $\ell$-independent, in the eikonal limit. We leave such an exploration for future work, but note that an eikonal expansion of the perturbative equations~\cite{Dolan:2009nk}, or working in the Penrose limit of the lightring~\cite{Fransen:2023eqj, Giataganas:2024hil, Kehagias:2024sgh, Kapec:2024lnr, Bucciotti:2025rxa, Kehagias:2025ntm, Fransen:2025cgv}, are possible paths to study such problem analytically. 

\begin{figure}
    \centering
    \includegraphics[width=\columnwidth]{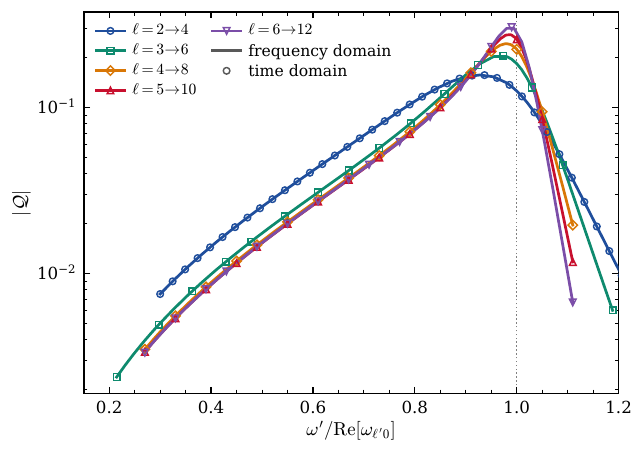}
    \caption{Self-coupling of GWs with angular harmonics $(
    \ell,\ell)$ driving the $2\ell$ harmonic, with frequency $\omega'$. The different colors correspond to different values of $\ell$. The empty markers are obtained from time-domain simulations, demonstrating the agreement of our two methods. We highlight that as $\ell$ grows the behavior appears universal, with a localized peak close to the resonance condition.  }
    \label{fig:selfcoup}
\end{figure}

\subsection{Geometry of $|\mathcal{Q}|$}

\begin{figure*}[t!]
    \centering
    \includegraphics[width=0.85\linewidth]{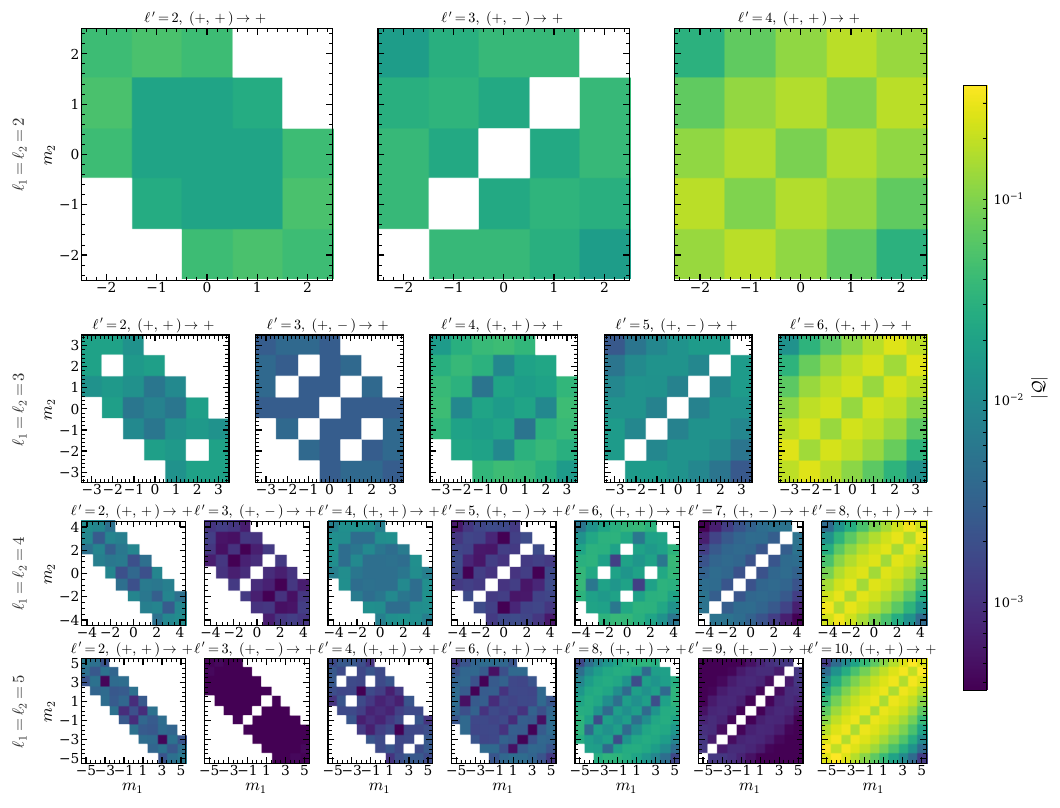}
    \caption{Dependence of the quadratic BH coupling coefficient $\mathcal{Q}$ with the azimuthal numbers $m_1,m_2$, for different values of $\ell_1=\ell_2=2,3,4,5$ in the different rows, and different target multipoles $\ell'$, in the different columns. We can see that the response of the BH is largest when $m_i = \ell_i = \ell'/2$, and that it is largely suppressed for some angular channels.}
    \label{fig:m_dependence}
\end{figure*}

A powerful feature of our approach is that it allows us to understand the dependence of the quadratic coupling coefficient on the geometry of the linear perturbations. In particular, given some fixed values of $\ell_1,\ell_2,\ell'$, the dependence on the azimuthal numbers $m_1$ and $m_2$~\footnote{Recall that $m'=m_1+m_2$ is fixed.} is known analytically in terms of $3j$-symbols. 
Writing the $m$-independent reduced radial source as $\widehat{\mathfrak S}$, the source is
\begin{equation}
    \begin{aligned}
    \mathfrak S_{\bs_1,\bs_2,\bs'}={}&(-1)^{m'}
    \begin{pmatrix}\ell_1&\ell_2&\ell'\\m_1&m_2&-m'\end{pmatrix}
    \widehat{\mathfrak S}_{\bs_1,\bs_2,\bs'} \, , 
    \end{aligned}
\end{equation}
where $\widehat{\mathfrak{S}}$ is $m$-independent. Therefore we can relate a particular $(m_1,m_2,m')$ channel given any nonvanishing reference triplet $(\bar m_1,\bar m_2,\bar m')$,
\begin{equation}
    \begin{aligned}
    \mathcal Q_{m_1m_2\to m'}={}&K_{m_1m_2m'}^{\bar m_1\bar m_2\bar m'}
    \mathcal Q_{\bar m_1\bar m_2\to\bar m'} \, ,\\
    K_{m_1m_2m'}^{\bar m_1\bar m_2\bar m'}={}&
    \frac{(-1)^{m'}\begin{pmatrix}\ell_1&\ell_2&\ell'\\m_1&m_2&-m'\end{pmatrix}}
    {(-1)^{\bar m'}\begin{pmatrix}\ell_1&\ell_2&\ell'\\\bar m_1&\bar m_2&-\bar m'\end{pmatrix}} \, .
    \end{aligned}
\end{equation}
The large angular number limit of this quantity was also studied e.g. in~\cite{Bucciotti:2025rxa}. Here we have tested directly this dependence, and we show it in Fig.~\ref{fig:m_dependence}. Moreover this figure also shows the dependence of $|\mathcal{Q}|$ on the final angular number $\ell'$. 

We notice two distinct features. First, the Wigner coefficient favors aligned and maximally co-rotating parent modes. The quadratic BH coupling coefficient is large for the maximal $\ell'=\ell_1+\ell_2$, and $m_i = \ell_i$. One could interpret this as a tendency of GWs to transfer energy efficiently to higher (angular) frequencies -- the frequency $\omega'=\omega_1+\omega_2=\omega_{\ell' 0}$ is always fixed to be the resonant frequency. We refrain from translating this finding into a statement about the potential turbulent dynamics of General Relativity, which deserves further scrutiny, see e.g.~\cite{Yang:2014tla,Green:2013zba, Ma:2025rnv, Lehner:2026tfe}. Secondly, we notice that this preferred transfer towards maximal angular momentum is not monotonic. For instance, the quadratic response towards the \emph{minimal} angular momentum ($\ell'=2$ in all cases considered here) is never the smallest. We do not show the $(5,m_1)\times (5,m_2) \to (\ell',m')$ coupling with $\ell'=5,7$, since the values we obtain are $\lesssim 10^{-4}$, and we cannot resolve them accurately enough. 

\subsection{Dependence on parity content}

\begin{figure}
    \centering
    \includegraphics[width=\columnwidth]{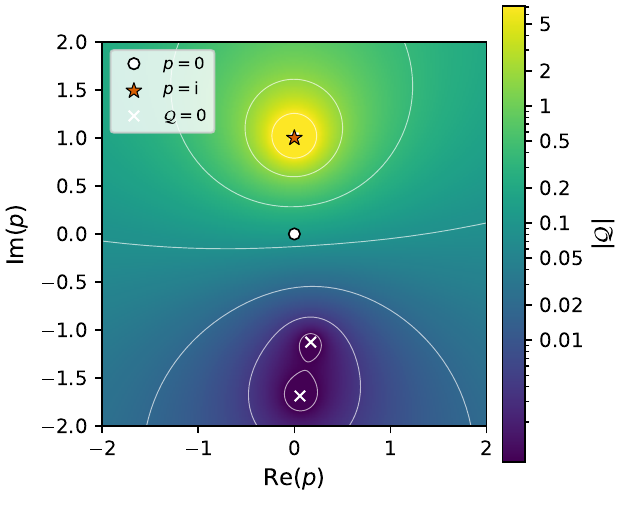}
    \caption{Dependence of the quadratic coupling coefficient $|\mathcal{Q}|$ as a function of the parity mixing $p = \psi^{(1)}_{-}/\psi^{(1)}_+$. The white dot corresponds to purely even parity initial data, and the star corresponds to the pole of $\mathcal{Q}$ due to zero incoming strain. White crosses denote zeros of $|\mathcal{Q}|$. }
    \label{fig:parity}
\end{figure}

Following Ref.~\cite{Bourg:2024jme} (see also~\cite{Khera:2024bjs, Bucciotti:2024jrv}), we can study the dependence on the parity content of the driving initial data of the quadratic BH coupling coefficient. Suppose the initial data is 
\begin{equation}
   \psi^{(1)}=C^+\psi^{(1)}_+ + C^-\psi^{(1)}_- \, ,
    \qquad p\equiv\frac{C^-}{C^+} \, .
\end{equation}
i.e., $p$ is the complex number which measures the ratio of odd to even parity perturbations in the driving GWs. For a given combination of angular harmonics $(\ell_i,m_i)$, and outgoing harmonic $\ell', m'=m_1+m_2$, not all parity combinations are admissible. However, considering the admissible parity combinations, we can define a total quadratic BH coupling coefficient, as a function of the parity mixing coefficient $p$. For instance, if the BH is driven with purely quadrupolar waves, $\ell_i=m_i=2$, then 
\begin{equation}
    |\mathcal Q(p)|=
    \frac{\left|\mathcal Q_{++\to+}+2ip\,\mathcal Q_{+-\to-}
    -p^2\mathcal Q_{--\to+}\right|}{|1+ip|^2}
\end{equation}
We have computed this quantity as a function of $p$ for the case of quadrupolar driving, and show its results in Fig.~\ref{fig:parity}. Note that $p=0$ corresponds to the case compatible with up-down reflection symmetry, whereas $p=i$ corresponds to zero incident strain, hence there is a pole in $|\mathcal{Q}|$. We find that the value of $\mathcal{Q}$ depends on the parity content, in particular near the pole $p=i$, and near $p=-i$, where we find two closely-located zeros, marked with white crosses. Near $p\sim 0$ the dependence of $|\mathcal{Q}|$ on the parity content is milder. 

\subsection{Driving by unequal frequencies}

\begin{figure*}
    \centering
    \includegraphics[width=0.9\linewidth]{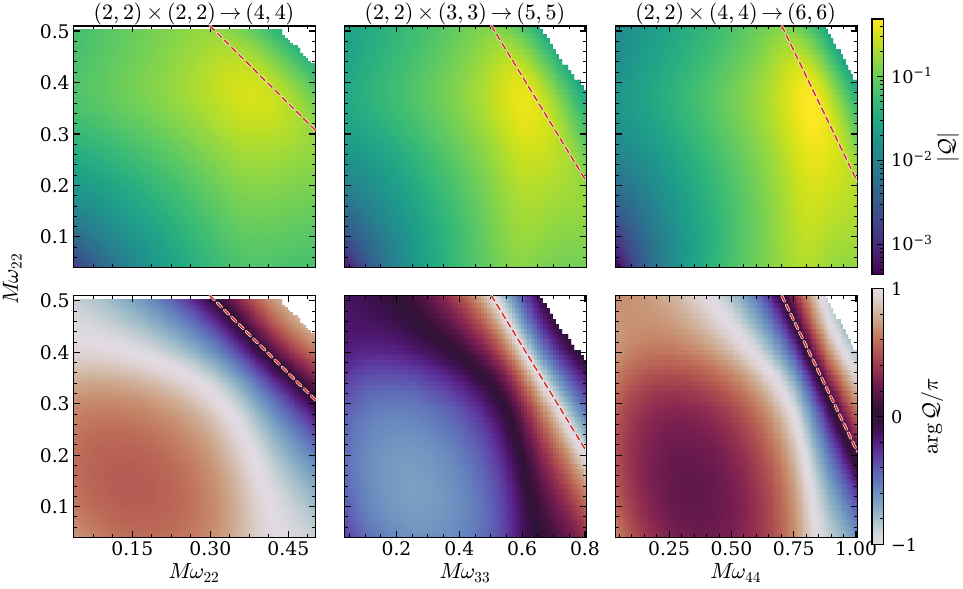}
    \caption{Absolute value (top row) and phase (bottom row) of $\mathcal{Q}_{++\to +}$ when driven by the $(2,2)$ mode combined with another $(2,2)$ mode (first column), with the $(3,3)$ mode (middle column), and with the $(4,4)$ mode (rightmost column). In all cases, we omit some high-frequency points, where our estimate for the convergence of the integral becomes larger than $1\%$. The dashed line represents the frequencies at resonance, i.e., $\omega_1 + \omega_2 = \Re[\omega_{\ell' m' 0}]$. 
    }
    \label{fig:combined}
\end{figure*}

Finally we study the behavior of $\mathcal{Q}$ when driving the BH with \emph{unequal} frequencies. Naively we expect that the system is only sensitive to $\omega'=\omega_1+\omega_2$. Fig.~\ref{fig:combined} studies this in detail. We have studied three different mode combinations, consisting of the $(2,2,+)$ mode coupling to itself, to the $(3,3,+)$, and to the $(4,4,+)$. In all cases the absolute value of $\mathcal{Q}$ peaks when $\omega'\sim \Re[\omega_{\ell' 0}]$. Nevertheless it is quite interesting that there is some nontrivial dependence on the individual frequencies. The peak of $\mathcal{Q}$ corresponds also to the point where each of the driving modes is close to its own QNM frequency. We note that at fixed $\omega'$, the value of $|\mathcal{Q}|$ changes significantly. The phase varies smoothly across frequencies -- seemingly matching $\arg\mathcal{Q} \sim 0$ near the resonant driving frequency.

\section{Scattering Amplitudes Perspective}

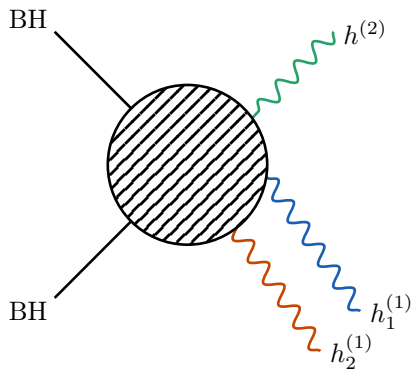
\begin{figure}
    \centering
    \input{feynman_diagram}
    \caption{$5$-point Compton scattering amplitude representing the nonlinear scattering of two gravitons $h^{(1)}_{1,2}$ into a single graviton $h^{(2)}$, mediated by the presence of a black hole represented here as a very massive (spinless) point particle with a particular gravitational coupling.}
    \label{fig:feynman}
\end{figure}

Our computation was performed using BH perturbation theory. As such it is nonperturbative in $G_N$ but treats the BH spacetime as a fixed background. Scattering amplitude techniques have recently been successfully applied to BH physics, accurately computing the post-Minkowskian 2 body Hamiltonian~\cite{Bern:2019crd, Bern:2021dqo}, scattering angle~\cite{Bjerrum-Bohr:2019kec, Bern:2019nnu, Bjerrum-Bohr:2021din, DiVecchia:2021bdo, Damour:2019lcq}, GWs emitted during scattering encounters~\cite{Herrmann:2021lqe, Herrmann:2021tct, Mougiakakos:2021ckm, DiVecchia:2022piu, Adamo:2022qci, Goldberger:2016iau}, and scattering of GWs off BHs~\cite{Bautista:2021wfy, Chiodaroli:2021eug, Bautista:2022wjf, CarrilloGonzalez:2025gqm}, to name some applications. Scattering techniques could offer a complementary perspective to the problem treated in this paper. 

Broadly speaking, amplitude computations are perturbative in $G_N$ and leverage the full Poincaré symmetry group of flat space, in addition to modern scattering amplitude techniques such as generalized unitarity. Nonlinear gravitational scattering of a black hole is schematically described by the amplitude shown in Fig.~\ref{fig:feynman},
\begin{equation}
    \begin{aligned}
            \mathcal{Q} &\sim \langle P', k', \varepsilon'| iT|P, k_1,\varepsilon_1,k_2,\varepsilon_2\rangle \, , \\ 
            &P+k_1+k_2 = P'+k' \, .
    \end{aligned}
\end{equation}
where $S=1+iT$ is the S--matrix, which encodes gravitational interactions and is computed perturbatively in $G_N$ through loop calculations, the in--state is two incoming GWs with momenta $k_i$ and helicity $\varepsilon_i$ interacting together with a BH (a heavy particle) with momenta $P$, and the out--state is the BH with momentum $P'$, in addition to the nonlinearly produced GW with momentum and helicity $k',\varepsilon'$. In this framework, the problem is under perturbative control only for waves at frequencies $G_N M \omega \ll 1$, corresponding to the left part of the diagram in Fig.~\ref{fig:low_freq}, where our BH perturbation theory computation should match the amplitude one. The amplitude computation, being fully analytic, could provide a sharp prediction of the low frequency behavior, where $\mathcal Q$ seems to be scaling quadratically with $\omega$. Below we summarize the main ingredients, and estimate the leading order behavior. A detailed calculation remains an interesting open problem.

Moreover, the differences between BH perturbation theory and scattering amplitudes calculations become apparent when advancing towards higher orders. We expect that the low-frequency behavior shown in e.g. Fig.~\ref{fig:low_freq} is well captured by the $5$-point Compton amplitude represented in Fig.~\ref{fig:feynman}. Indeed, below we will show that the $\mathcal Q\sim \omega^2$ scaling can easily be reproduced from it. Going to higher orders in $G_N M \omega$ is challenging from the scattering amplitudes point of view, as it involves loop calculations, but is straightforward from BH perturbation theory. On the other hand, including an additional GW (graviton) remains an open problem in BH perturbation theory, since one needs to compute third-order perturbations -- currently an open problem. A leading order (in $G_N M \omega$) result may be more easily accessible from the scattering amplitudes point of view, by computing a six-point graviton Compton amplitude. 

Let us estimate the scattering amplitudes result, to leading order, for the quadratic BH response studied in this work. We will show that the 5--point Compton amplitude for three gravitons and a nonrelativistic heavy object reproduces the $\omega^2$ scaling in the left part of Fig.~\ref{fig:low_freq}. This amplitude was computed in~\cite{Bjerrum-Bohr:2023jau}, and is given by
\begin{align}
    \mathcal{M}_5 = {[\mathcal{N}_0([1,[2,3]],v)]^2\over p_{23}^2\,p_{123}^2}
+{[\mathcal{N}_0([[1,2],3],v)]^2\over p_{12}^2\,p_{123}^2}+\\[5pt]
+{[\mathcal{N}_0([[1,3],2],v)]^2\over p_{13}^2\,p_{123}^2}  \, ,
\end{align}
where we specialized to a non-spinning black hole, and the $p$'s are various combinations of the graviton momenta, thus proportional to $\omega$. From the definition of $\mathcal{N}_0$, it is easy to extract the scaling with $M$ and $\omega$. Letting $F_i^{\mu\nu} = p_i^\mu\varepsilon_i^\nu - p_i^\nu\varepsilon_i^\mu \sim \omega$,
\begin{equation}
    \mathcal{N}_0 \sim M \frac{F^3}{p} \sim M \omega^2 \, \Rightarrow \,
    \mathcal{M}_5 \sim \kappa^3 M^2 \omega^0 \, ,
\end{equation}
where one factor $\kappa$ ($\kappa^2=32\pi G_N$) appears for each graviton. To obtain the waveform from the amplitude, we apply the techniques in \cite{Cristofoli:2021vyo}, which schematically give
\begin{align}
    \dd\Phi_k &= \frac{\dd^3 k}{(2\pi)^3 2\omega_k} \sim \omega^2  \, , \\
    h_{\mu\nu} &= \kappa \sum_{\lambda=1,2} \int \dd \Phi_k\, \varepsilon_{\mu\nu}^\lambda(\vec k) \alpha^\lambda(\vec k) e^{-ikx} \; +\, h.c.  \, \sim \kappa \omega^2 \alpha  \, , \\
    \alpha^{\rm out} &\simeq \frac{1}{2M} \sum_{\lambda_1,\lambda_2} \int \dd \Phi_{k_1} \dd \Phi_{k_2} \delta(\vec k_1+\vec k_2+\vec k_3) \mathcal{M}_5\, \alpha^{\rm in}_1 \alpha^{\rm in}_2  \, ,
\end{align}
where the crucial factor $1/M$ comes from the nonrelativistic limit of $\frac{1}{2\omega_M}$ where $\omega_M$ is the energy of the black hole. Noting that $\alpha^{\rm out}\sim \kappa^3 M \omega^3 (\alpha^{\rm in})^2$, one discovers that
\begin{equation}
    h^{\rm out} \sim \kappa^2 M \omega (h^{\rm in})^2  \, ,
\end{equation}
and recalling eq.~\ref{eq:strain_amplitude} which schematically gives $h\sim A/r\sim A\omega$, the final result is
\begin{equation}
    A^{\rm out}\sim G_N M\omega^2 (A^{\rm in})^2  \, ,
\end{equation}
in perfect agreement with Fig.~\ref{fig:low_freq}. Up to numerical coefficients, the result is independent of angular momenta, which only enter in the precise relation between $h_{\mu\nu}$ and the amplitudes, and is independent of the parities, that are specified by the $\varepsilon_{\mu\nu}$ tensors. Using these techniques and keeping track of all numerical factors, it should be possible to also derive the coefficient of the power-law as well. We leave this as an open problem.

Our conclusion is that nonlinear effects appear as a new arena where BH perturbation theory and scattering amplitude techniques can complement each other. As an example, only in the former framework can we reliably push to frequencies comparable to (the real part of) QNMs, and see the resonances shown in Fig.~\ref{fig:resonance}, \ref{fig:selfcoup}. On the other hand the low frequency expansion encodes nonlinear (quadratic) dynamical Love numbers, which represent the tidal and dissipative nonlinear response of an object when perturbed by an external field, at frequencies below the excitation gap. 

\section{Conclusions}

The linear response of a BH to gravitational perturbations has been understood for a long time. It encodes key observable effects in BH physics such as the static and dynamic response of the BH to external gravitational fields (Love numbers), greybody factors, and, through analytical continuation, the free oscillations of the BH, i.e., QNMs. 

In this work we have computed for the first time the quadratic coupling coefficient of a non-rotating BH to gravitational perturbations, for generic parity, frequency, and angular harmonic combinations. This calculation not only furthers our understanding of BH perturbation theory beyond linear order, but captures some major physical insights in the nonlinear regime, including nonlinear dynamical BH tidal deformabilities and resonances when driving a BH near its QNM frequency. The former may be imprinted in subtle ways in the GW dephasing during binary BH mergers, while the latter can be enhanced in the presence of e.g. hierarchical triple systems~\cite{Cardoso:2021vjq,Santos:2025ass, Santos:2026lzq}. We highlight that we implement two different methods: in the time-domain, and in the frequency-domain, and that the findings of this work are in excellent agreement with the numerical findings of~\cite{Keeble:2026epf}, obtained from nonlinear experiments scattering purely quadrupolar wave packets (albeit in a different normalization). 

One of our major results is demonstrating the quadratic scaling of the BH response with the driving frequency. This (numerical) finding is further justified by arguments that start from the $5$-point Compton scattering amplitude for gravitons~\cite{Bjerrum-Bohr:2023jau}, showing agreement between amplitude techniques and our calculation. A very interesting open direction is to further this match: can scattering amplitudes calculations reproduce the quadratic BH coupling coefficient beyond linear order? More interestingly, it is likely that scattering amplitudes techniques may be better suited than BH perturbation theory to go to higher orders in perturbation theory. 

A natural new direction is studying the same process for rotating BHs. While conceptually the problem remains unchanged, we note that a fundamental difficulty is that the first order metric is most often reconstructed in a radiation gauge, which is not suitable for scattering data such as the one prescribed here. Lorenz gauge metric perturbations may lead to better behaved sources for the second-order Teukolsky equation with scattering data which, nevertheless, may require further regularization at infinity and at the event horizon. This is an exciting open direction, in particular in regards to the behavior of the quadratic BH coupling coefficient in the near-extremal limit, which we leave for future work.

\begin{acknowledgments}
B.B. and A.K. thank the Center of Gravity at the Niels Bohr Institute for hospitality while part of this work was being carried out.
We thank Zvi Bern, Fei Teng, and Maarten van de Meent for discussions. We are grateful to Lennox Keeble and Hengrui Zhu for facilitating a comparison with their nonlinear results. We also acknowledge valuable discussions on the connection between BH perturbation theory and scattering amplitudes with Nabha Shah, Emil Bjerrum-Bohr, and Julio Parra-Martínez. 
B.B. is supported by the U.S. Department of Energy under grant number DESC0019470 and by the Heising-Simons Foundation `Observational Signatures of Quantum Gravity' collaboration grant 2024-5305.
A.K. thanks the Fundação para a Ciência e Tecnologia (FCT), Portugal, for the financial support to the Center for Astrophysics and Gravitation (CENTRA/IST/ULisboa) through grant No. UID/PRR/00099/2025 and grant No. UID/00099/2025, as well as to the FCT project ``Gravitational waves as a new probe of fundamental physics and astrophysics'' grant agreement 2023.07357.CEECIND/CP2830/CT0003.
The Center of Gravity is a Center of Excellence funded by the Danish National Research Foundation under grant No. DNRF184.
We acknowledge support by VILLUM Foundation (grant no. VIL37766).
V.C.\ is a Villum Investigator.  
V.C. acknowledges financial support provided under the European Union’s H2020 ERC Advanced Grant “Black holes: gravitational engines of discovery” grant agreement no. Gravitas–101052587. 
Views and opinions expressed are however those of the author only and do not necessarily reflect those of the European Union or the European Research Council. Neither the European Union nor the granting authority can be held responsible for them.
This project has received funding from the European Union's Horizon 2020 research and innovation programme under the Marie Sklodowska-Curie grant agreement No 101007855 and No 101131233.
This work is supported by Simons Foundation International \cite{sfi} and the Simons Foundation \cite{sf} through Simons Foundation grant SFI-MPS-BH-00012593-11.
Z.Z. acknowledges financial support from the MUR FIS2 Advanced Grant ET-NOW (CUP:~B53C25001080001) and the INFN TEONGRAV initiative.
\end{acknowledgments}
\appendix

\section{Accuracy and Convergence}

\begin{figure}[h!]
    \centering
    \includegraphics[width=\columnwidth]{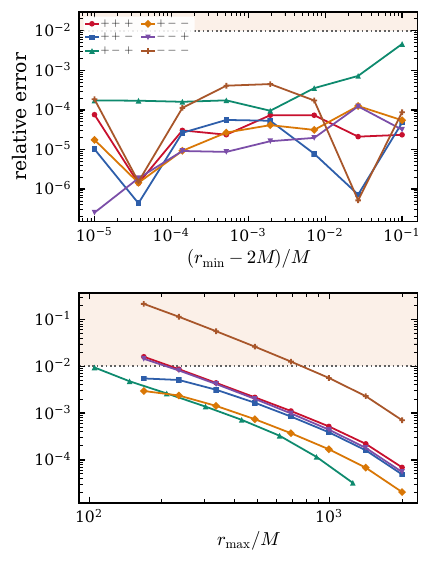}
    \caption{Relative error in $\mathcal{Q}$ when varying the inner (top) and outer (bottom) integration limits. We push the integration limits until the error in $\mathcal{Q}$ is sub-percent. }
    \label{fig:convergence}
\end{figure}

Here we detail our tests of the accuracy and convergence of our numerical calculation of $\mathcal{Q}$. For every point computed, including all values shown in the Figures in the main text, we test that $\mathcal{Q}$ does not depend if we change either limit of integration, and that the final value of $\mathcal{Q}$ does not vary if we add an extra piece to the regularization functions $\Delta_i$, in a way that it does not alter the asymptotic structure. The first is simply a test of the accuracy of our numerical solvers of the differential equations, and the numerical integration itself. The former, on the other hand, is a nontrivial test that the source regularization is insensitive to the addition of higher-order counterterms. We vary these three factors (inner and outer integration limits, and addition of an extra free coefficient), and define the error $\epsilon$ as the maximum relative error in the calculation of $\mathcal{Q}$ when varying these three factors. We discard any point where $\epsilon > 10^{-2}$. 

It is illustrative to show the convergence of $\mathcal{Q}$ in particular when varying the integration limits. This is shown in Fig.~\ref{fig:convergence}, for different parity combinations. As expected, the integral is most sensitive to the outer integration limit -- despite including Richardson extrapolation to remove the $1/r$ tail, the source only decays as $1/r^2$, which makes this slowly converging. However, we find that the errors decrease steadily and we can achieve enough accuracy for our purposes. We also note that one could add additional counterterms to the functions $\Delta_i$ and the second-order gauge vector, in order to make the effective source $\mathfrak{S}$ decay faster than $r^{-2}$. This would potentially facilitate a numerical analysis of the behavior of $|\mathcal{Q}|$ as $G_N M \omega \to \infty$, which is the most challenging regime of our current numerical implementation.

\bibliography{biblio}

\clearpage 
\onecolumngrid

\end{document}

%% file: feynman_diagram.tex
\definecolor{fdgcol0}{HTML}{000000}
\definecolor{fdgcol1}{HTML}{C64600}
\definecolor{fdgcol2}{HTML}{1A5FB4}
\definecolor{fdgcol3}{HTML}{26A269}
\definecolor{fdgcol4}{HTML}{FFFFFF}
\begin{tikzpicture}
  \path[draw=fdgcol0, line width=1pt] (75pt,-150pt) -- (125pt,-100pt);
  \path[draw=fdgcol0, line width=1pt] (75pt,-50pt) -- (125pt,-100pt);
  \path[draw=fdgcol1, line width=1pt, decorate, decoration={snake, amplitude=3pt, segment length=10pt, pre length=0pt, post length=0pt}] (175pt,-170pt) -- (138pt,-117pt);
  \path[draw=fdgcol2, line width=1pt, decorate, decoration={snake, amplitude=3pt, segment length=10pt, pre length=0pt, post length=0pt}] (190pt,-155pt) -- (153pt,-102pt);
  \path[draw=fdgcol3, line width=1pt, decorate, decoration={snake, amplitude=3pt, segment length=10pt, pre length=0pt, post length=0pt}] (180pt,-50pt) -- (149.333pt,-82.452pt);
  \path[fill=fdgcol4] (125pt,-100pt) ellipse [x radius=30pt, y radius=30pt];
  \path[pattern={Lines[angle=45, distance=4pt, line width=1pt]}, pattern color=fdgcol0] (125pt,-100pt) ellipse [x radius=30pt, y radius=30pt];
  \path[draw=fdgcol0, line width=1pt] (125pt,-100pt) ellipse [x radius=30pt, y radius=30pt];
  \node[inner sep=0pt] at (65pt,-155pt) {BH};
  \node[inner sep=0pt] at (65pt,-45pt) {BH};
  \node[inner sep=0pt] at (202pt,-155pt) {$h_1^{(1)}$};
  \node[inner sep=0pt] at (187pt,-170pt) {$h_2^{(1)}$};
  \node[inner sep=0pt] at (192pt,-50pt) {$h^{(2)}$};
\end{tikzpicture}